\documentclass{article}
\usepackage[a4paper, total={5.5in, 9.5in}]{geometry}
\usepackage{siunitx}
\usepackage{bm}
\usepackage{amsmath}
\usepackage{amsfonts}
\usepackage{physics}
\usepackage{graphicx}
\usepackage{authblk}
\usepackage{natbib}
\usepackage{xcolor}
\usepackage{lineno}
\usepackage{amssymb}
\usepackage{algorithm}
\usepackage{algorithmic}
\usepackage{booktabs}
\usepackage{setspace}

\title{A numerical stabilization scheme for the shallow shelf approximation}
\author[1]{Tilda Westling Dolling}
\author[1,2]{A. Clara J. Henry}
\author[1,2,3]{Josefin Ahlkrona}

\affil[1]{Department of Mathematics, Stockholm University, Stockholm, Sweden}
\affil[2]{Bolin Centre for Climate Research, Stockholm University, Stockholm, Sweden}
\affil[3]{Swedish e-Science Research Centre, Stockholm, Sweden}

\date{}

\begin{document}
\setstretch{1.1}

\maketitle

\noindent\rule{\linewidth}{0.4pt}

\noindent
\textbf{Correspondence:} A. Clara J. Henry (clara.henry@math.su.se)

\subsection*{Key points:}
\begin{itemize}
\item We introduce the Thickness Stabilization Scheme (TSS), developed for vertically-integrated ice-sheet models and designed to mimic an implicit treatment of the driving force.
\item The TSS allows significantly larger stable time steps and, in particular, enables time steps at least twice as large to remain within 1\% accuracy.
\end{itemize}

\noindent\rule{\linewidth}{0.4pt}

\section*{Abstract}

We present the Thickness Stabilization Scheme (TSS), a numerical stabilization scheme suitable for the Shallow Shelf Approximation (SSA), one of the most widely-used models for large-scale Antarctic and Greenland ice sheet simulations. The TSS is constructed by inserting an adapted, explicit Euler thickness evolution equation into the driving stress term, thereby treating the term implicitly. We investigate the applicability of TSS across low- and high-shear idealized scenarios, by altering the inflow velocity and initial ice thickness. TSS demonstrates an increase in the numerical stability of SSA, allowing large time-step sizes of $\Delta t = 50-100$ years to remain numerically stable and accurate, while time-step sizes of $\Delta t > 5$ in high-shear simulations show significant error without TSS. Remarkably, a time step size as large as $\Delta t = 10\,000$ years is numerically stable with TSS, albeit with a reduction in accuracy. TSS offers greater flexibility for ice-sheet modeling by allowing the re-allocation of computational resources. This method is applicable not only to ice-sheet modeling, including in coupled frameworks, but also to other vertically-integrated computational fluid dynamics problems that couple momentum and geometry evolution equations.

\section{Introduction}

The efficiency and accuracy of numerical ice-sheet models continues to be a challenge, with model intercomparison projects showing large model variability \cite{Seroussi2024}. A key contributing factor is the need for small time steps to ensure numerical stability. For example, it has been reported across five frequently used, lower-order ice-flow models that time-step sizes of $\Delta t \geq 5$ years lead to numerical instability in an idealized setting \cite{Robinson2022} and a high computational cost when simulating large domains such as the Antarctic Ice Sheet. Ice flow is typically described by a system of coupled partial differential equations (PDEs), which is solved using methods such as the finite element method (FEM) \cite{Gagliardini2013, Larsen2013}. This system of equations involves coupling a momentum equation with equations that allow evolution of the geometry such as a free-surface or thickness evolution equation. The coupling between the geometry and velocity in ice-sheet models, as well as similar systems, hinders the applicability of fully implicit time-stepping schemes, which generally offer greater numerical stability and larger time-step sizes.

To address numerical instability and accuracy challenges in such coupled models, various techniques have been used, including artificial viscosity, adaptive mesh refinement, and specialized solvers for large systems of equations \cite{Wilkins1980,Cornford2016,Larour2012}. The issue of a restrictive time step has been addressed in p-Stokes or full Stokes models coupled with free-surface evolution through stabilization techniques such as the sea spring method \cite{Durand2009} and the free-surface stabilization algorithm (FSSA) \cite{Kaus2010, Loefgren2022, Loefgren2024, Henry2025FSSA}. The sea spring stabilization scheme predicts the surface elevation of the ice-ocean interface in the next time step to treat the ocean pressure boundary condition implicitly. In contrast, the FSSA predicts the surface elevation of the free surface(s) in the next time step to treat the driving force implicitly. Such numerical stabilization schemes increase the largest stable time-step size by 1-2 orders of magnitude, depending on the application \cite{Loefgren2022,Loefgren2024,Henry2025FSSA,Ahlkrona2025}. Although such numerical advances significantly increase the applicability of p-Stokes models to areas of complex flow, large parts of ice sheets do not require the complexity of such a high-fidelity model.

Lower-order, depth-integrated models are computationally more efficient than p-Stokes models, but can nonetheless suffer from a restrictive time step. For example, Robinson et al., 2022 \cite{Robinson2022} report that neither the Shallow Shelf Approximation (SSA) nor any other depth-integrated ice-flow model investigated was numerically-stable for time-step sizes of $\Delta t > 5$ years. In this study, we focus specifically on the SSA, which is a 2D, vertically-integrated, depth-independent ice-flow model, suitable for application in ice shelves and ice streams, and is one of the most widely-used models for large-scale Antarctic Ice Sheet simulations, such as in the ISMIP6 model intercomparison \cite{Seroussi2024}. The SSA is a coupled system of PDEs, which consists of a set of time-independent, horizontal momentum equations and a thickness evolution equation. In this coupled framework, the horizontal velocity solution of the momentum equations enters as coefficients in the thickness evolution equation. Conversely, the solution of the thickness evolution equation enters into the momentum equation in the right-hand side forcing term. However, unlike p-Stokes formulations, the SSA does not include a vertical velocity component, making the implementation of FSSA nontrivial, as it relies on a vertical velocity component. Recognizing that the FSSA increases the largest stable time step of p-Stokes simulations by treating the right-hand side of the p-Stokes momentum equations implicitly, we construct a numerical scheme that similarly treats the right-hand side of the SSA momentum equations implicitly.

\begin{figure}
    \centering
    \includegraphics[width=0.7\linewidth]{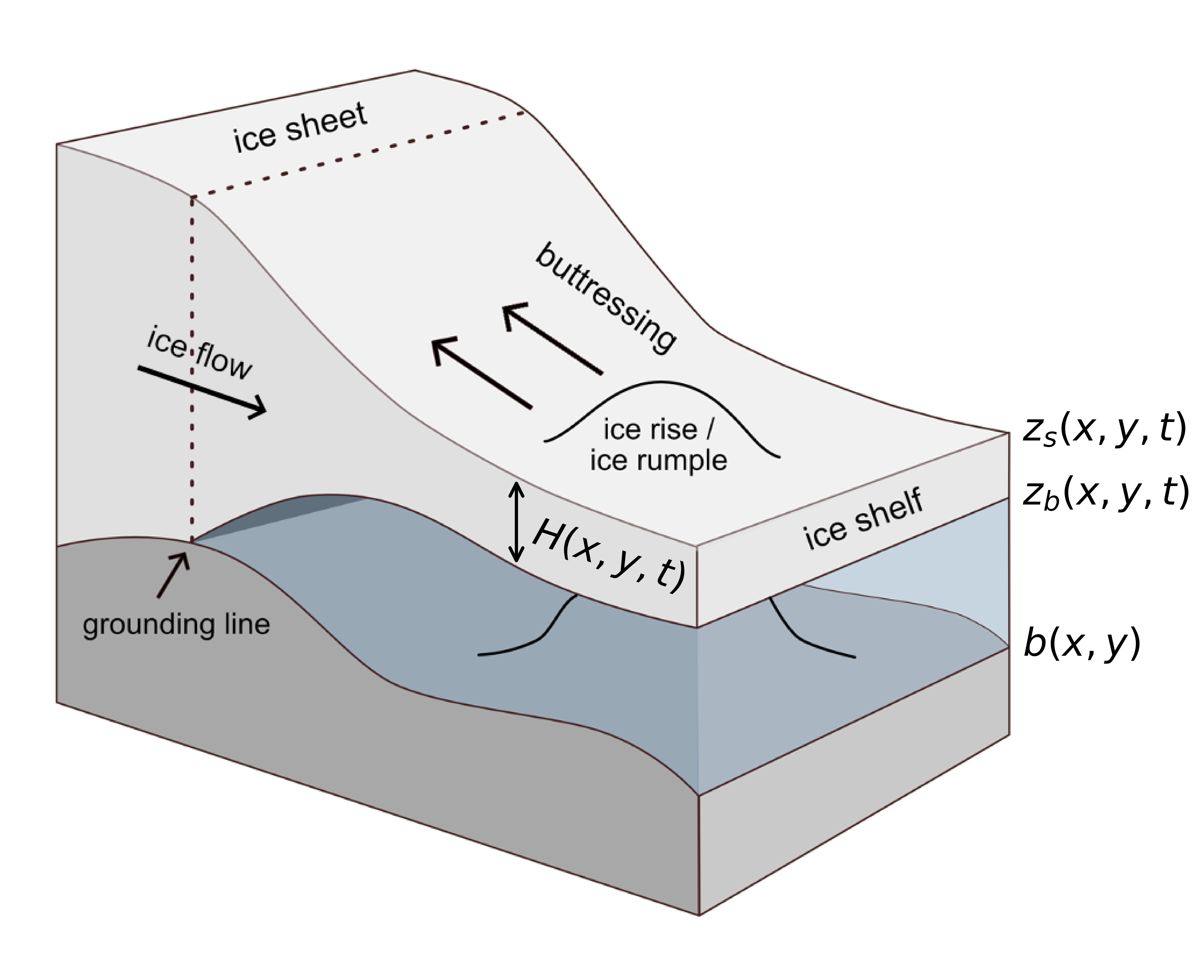}
    \caption{The flow of ice in a marine-terminating domain. The upper ice surface, $z_s(x, y, t)$, represents the ice-atmosphere interface, the lower ice surface, $z_b(x, y, t)$, represents the ice-bed or ice-ocean interface, and $b(x, y)$ represents the bed elevation. The ice sheet flows with contact to the bedrock below ($z_b = b$) before beginning to float ($z_b > b$) at the grounding line to form an ice shelf. The otherwise floating ice can come into contact with anomalies in the bedrock to form pinning points referred to as ice rises or ice rumples, where $z_b = b$. The ice thickness is the vertical distance between the upper and lower ice surfaces, i.e. $H(x, y, t) = z_s(x, y, t) - z_b(x, y, t)$.}
    \label{fig:ice_shelf}
\end{figure}

In this study, we introduce the thickness stabilization scheme (TSS) for the SSA model. Where FSSA predicts the free-surface elevation at the next time step, the TSS instead predicts the ice thickness using an adapted explicit Euler thickness evolution equation to replace the right-hand driving force. We investigate the computational efficiency and accuracy of SSA, with and without TSS, in simulations of floating ice flowing past a cylindrical obstacle. Results are presented and evaluated for simulations with varying initial ice thickness conditions and time-step sizes. Simulations with TSS remain numerically-stable and accurate with time-step sizes of $\Delta t = 50-100$ years, whereas time-step sizes of $\Delta t > 5$ in high-shear simulations show significant error without TSS. This work enables greater efficiency in ice-shelf simulations, and has the potential to be included in other coupled finite element modeling frameworks \citep{Seroussi2012, Ahlkrona2016}. The applicability of TSS reaches beyond ice-sheet modeling, with the potential to increase the numerical stability of other shallow fluid models more generally.

\section{Governing equations}

The SSA is an approximation derived from the Stokes equations,
\begin{align} \label{eq:stokes}
    \nabla \cdot \boldsymbol{\sigma}&= \rho_i \mathbf{g} \quad \text{in} \,\Lambda(x,y,z, t),\\
    \nabla \cdot \mathbf{u} &= 0 \quad \text{in} \,\Lambda(x,y,z, t),
\end{align}
where $\mathbf{u}_\mathrm{3D}=(u_x,u_y,u_z)$ is the 3D velocity vector, $\rho_i$ is the density of ice, $\mathbf{g}=(0,0,-g)$ is the gravitational acceleration, and $\Lambda(x, y, z, t)$. The Cauchy stress tensor, $\boldsymbol{\sigma} = \boldsymbol{\tau} - p\mathbf{I}$,  depends on the deviatoric stress, $\boldsymbol{\tau}$, and the pressure, $p$. 
 A nonlinear constitutive relationship, called Glen's flow law, relates the deviatoric stress tensor to the strain rate tensor,
\begin{align}\label{tau}
    \boldsymbol{\tau}= 2 \eta(\dot{\boldsymbol{\varepsilon}} (\mathbf{u}_\mathrm{3D})) \dot{\boldsymbol{\varepsilon}} (\mathbf{u}_\mathrm{3D}),   \text{with} \quad  \eta = \frac{1}{2} A^{-1/n} \left( \frac{1}{2}\|\dot{\boldsymbol{\varepsilon}}\|^2_F + \dot{\varepsilon}_0^2 \right)^{(1-n)/(2n)},
\end{align}
where the strain rate is the symmetric part of the velocity gradient tensor, $\dot{\boldsymbol{\varepsilon}} (\mathbf{u}_\mathrm{3D})=\frac{1}{2}(\nabla \mathbf{u}_\mathrm{3D} + (\nabla \mathbf{u}_\mathrm{3D})^{\top})$, $A$  is the ice fluidity, $n$ is the Glen's flow law exponent and $\dot{\varepsilon}_0$ is a regularization parameter to prevent singularities at very low strain rates. Substituting the nonlinear constitutive relationship (\ref{tau}) into the Stokes equations (\ref{eq:stokes}) yields the p-Stokes equations, with $p = (n+1)/n$ (e.g., \cite{Hirn2013}).

In comparison to the p-Stokes equations, the SSA equations neglect vertical shear stresses,  $\sigma_{xz}$ and $\sigma_{yz}$, so that the $z$-component of \eqref{eq:stokes} becomes
\begin{align}
    \frac{\partial \sigma_{zz}}{\partial z} = \rho_i g,
\end{align}
which, when integrated vertically, provides an expression for the pressure,
\begin{align} \label{eq:hydrop}
    p = \rho_ig(z_s-z) + \tau_{zz}= \rho_ig(z_s-z) - \tau_{xx} - \tau_{yy},
\end{align}
where $z_s$ is the upper ice surface.  Inserting this expression into the $x$- and $y$-component of \eqref{eq:stokes} results in
\begin{subequations}
\begin{align}
    &2\frac{\partial\tau_{xx}}{\partial x}
    + \frac{\partial\tau_{yy}}{\partial x}
    + \frac{\partial \sigma_{xy}}{\partial y} 
    = \rho_ig\frac{\partial z_s}{\partial x}  \quad \text{in} \,\Lambda(x,y,z, t),\\ 
    &2\frac{\partial \tau_{yy}}{\partial y} 
    + \frac{\partial \tau_{xx}}{\partial y}
    + \frac{\partial \sigma_{xy}}{\partial x}
    = \rho_ig\frac{\partial z_s}{\partial y}  \quad \text{in} \,\Lambda(x,y,z,t).
\end{align}
\end{subequations}

Vertically integrating and using (\ref{tau}) to replace the deviatoric stress tensor components with the strain rates, the SSA momentum equations read
\begin{subequations} \label{eq:SSAfull}
\begin{align} 
    &4 \frac{\partial}{\partial x}\biggl(\bar\eta \frac{\partial u_x}{\partial x} \biggl) + 2\frac{\partial}{\partial x}\biggl( \bar\eta \frac{\partial u_y}{\partial y}  \biggl)
    +\frac{\partial}{\partial y}\biggl(\bar\eta \biggl(\frac{\partial u_x}{\partial y} 
    + \frac{\partial u_y}{\partial x} \biggl)\biggl) 
    =\rho_i g H\frac{\partial z_s}{\partial x}  \quad \text{in} \, \Omega(x,y) , \label{eq:SSAfullx} \\
    &4 \frac{\partial}{\partial y}\biggl(\bar\eta \frac{\partial u_y}{\partial y} \biggl) + 2\frac{\partial}{\partial y}\biggl( \bar\eta \frac{\partial u_x}{\partial x}  \biggl)
    + \frac{\partial}{\partial x}\biggl(\bar\eta \biggl(\frac{\partial u_x}{\partial y} 
    + \frac{\partial u_y}{\partial x} \biggl)\biggl) 
    = \rho_i g H\frac{\partial z_s}{\partial y}\quad \text{in} \, \Omega(x,y) , \label{eq:SSAfully}
\end{align}
\end{subequations}
where $\mathbf{u} = (u_x, u_y)$ is the horizontal velocity vector, $z_b$ is the lower ice surface, $H=z_s-z_b$ is the ice thickness, and $\bar\eta = \eta H$ is the vertically-integrated viscosity. As a result, the degrees of freedom are reduced from three velocity components and pressure defined in the three-dimensional domain, $\Lambda(x,y,z, t)$, to only two horizontal velocity components defined in the two-dimensional domain $\Omega(x,y,t)$.

In this study, we assume that the ice load and the ocean pressure are in balance so that the upper ice surface is directly related to the ice thickness, the ice density and the ocean density. This hydrostatic equilibrium assumption takes the form
\begin{align}
    z_s = \left( \frac{\rho_o - \rho_i}{\rho_o} \right)H,
\end{align}
where $\rho_o$ is the ocean density. The right-hand side of ~(\ref{eq:SSAfull}) can be altered so that
\begin{subequations}
\begin{align}
    4 \frac{\partial}{\partial x}\biggl(\bar\eta \frac{\partial u_x}{\partial x} \biggl) 
    + 2\frac{\partial}{\partial x}\biggl( \bar\eta \frac{\partial u_y}{\partial y}  \biggl)
    + \frac{\partial}{\partial y}\biggl(\bar\eta \biggl(\frac{\partial u_x}{\partial y} + \frac{\partial u_y}{\partial x} \biggl)\biggl) = \frac{1}{2}\rho_i g  \biggl(1-\frac{\rho_i}{\rho_o} \biggl)\frac{\partial H^2}{\partial x},
    \\
    4 \frac{\partial}{\partial y}\biggl(\bar\eta \frac{\partial u_y}{\partial y} \biggl) 
    + 2\frac{\partial}{\partial y}\biggl( \bar\eta \frac{\partial u_x}{\partial x}  \biggl)
    + \frac{\partial}{\partial x}\biggl(\bar\eta \biggl(\frac{\partial u_x}{\partial y} + \frac{\partial u_y}{\partial x} \biggl)\biggl) = \frac{1}{2}\rho_i g  \biggl(1-\frac{\rho_i}{\rho_o} \biggl)\frac{\partial H^2}{\partial y}.
\end{align}
\end{subequations}
By defining the tensor
\begin{equation}
    \mathbf{T} = \begin{pmatrix} \bar\eta  \big( 4 \frac{\partial u_x}{\partial x} + 2 \frac{\partial u_y}{\partial y} \big) & \bar\eta  \big( \frac{\partial u_x}{\partial y} + \frac{\partial u_y}{\partial x} \big) \\ \bar\eta  \big( \frac{\partial u_x}{\partial y} + \frac{\partial u_y}{\partial x} \big) & \bar\eta  \big( 2 \frac{\partial u_x}{\partial x} + 4 \frac{\partial u_y}{\partial y} \big) \end{pmatrix},
\end{equation}
and using a scaled gravitational term,
\begin{equation}
    \rho' = \frac{1}{2}\rho_i g \bigg(1-\frac{\rho_i}{\rho_o} \bigg),
\end{equation}
the SSA momentum equations can be rewritten as
\begin{equation} \label{eq:compactSSA}
    \nabla \cdot \mathbf{T} = \rho' \nabla H^2.
\end{equation}
This problem is subject to boundary conditions and the specific conditions used in this study are stated in Section \ref{sq:setup}.

\subsection{The thickness evolution equation} \label{sec:thick}

The geometry of the ice shelf evolves subject to a thickness evolution equation,
\begin{align}
    \frac{\partial H}{\partial t} &= - \nabla \cdot (H \mathbf{u}) + a_s - a_b,
\end{align}
where $a_s$ is the surface accumulation rate and $a_b$ is the melt rate. The equation is subject to a minimum ice thickness constraint $H\geq 10$ m, which is enforced after each thickness solve. Since the velocity enters as coefficients in the thickness evolution equation and the thickness enters the driving force in the SSA, the two equations are strongly coupled.

The thickness evolution equation is typically discretized with a first-order scheme such as the forward Euler method,
\begin{align} \label{eq:forwardH}
    H_{k+1} =H_k - \Delta t \nabla \cdot (H_k \mathbf{u}_k) + \Delta t (a_s - a_b),
\end{align}
where $H_k$ is the ice thickness at the current time step and $H_{k+1}$ is the ice thickness at the next time step. Alternatively, a semi-implicit method backward Euler scheme can be used,
\begin{align} \label{eq:backwardH}
    H_{k+1} + \Delta t \nabla \cdot (H_{k+1} \mathbf{u}_k) = H_k + \Delta t (a_s - a_b).
\end{align}
Coupling the momentum and geometry evolution equations using time-stepping schemes that are implicit in both the velocity and thickness are not used in ice sheet modeling since such a system requires computing  $\mathbf{u}_{k+1}$, which is notoriously difficult due to the coupling between the SSA and the thickness evolution equation. Since the coupled ice-sheet models are stiff \cite{Bueler2024}, the schemes of \eqref{eq:backwardH} and \eqref{eq:forwardH} are only conditionally stable, i.e. there is a restriction on the time-step size. The forward Euler time-stepping discretization (\ref{eq:forwardH}) is central to this study as it is used to construct the TSS, (\ref{TSS}). However, the semi-implicit backward Euler scheme (\ref{eq:backwardH}) will be used to integrate the thickness evolution in time.

\subsection{The variational form}

To solve the p-Stokes or the SSA model using the finite element method, it is necessary to rewrite the equations in weak form. The variational form of the SSA momentum equations is, find  $\mathbf{u}\in \boldsymbol{\chi} $ so that 
\begin{equation}\label{weak_no_boundaries}
    \int_{\Omega}\mathbf{T}(\mathbf{u}) : \nabla \mathbf{v} ~d\Omega = \rho' \int_{\Omega} H^2 \nabla\cdot \mathbf{v} ~d\Omega \quad \forall  \, \mathbf{v} \in \boldsymbol{\chi},
\end{equation}
where $\mathbf{v} \in \boldsymbol{\chi}$ is a test function living in an appropriate Sobolev space. The variational form is constructed by multiplying by the test function, integrating by parts and applying boundary conditions so that the boundary terms cancel. Examples of such boundary conditions are found in Section \ref{sq:setup}.

In weak form, the semi-implicit thickness evolution equation \eqref{eq:backwardH} reads
\begin{align} \label{eq:Hweak}
    \int_{\Omega} [ H_{k+1} + \Delta t \nabla \cdot (H_{k+1}\mathbf{u}) ] \phi \, d\Omega
= \int_{\Omega} [ H_k + \Delta t (a_s - a_b ) ] \phi \, d\Omega,
\end{align}
where $\phi$ is a test function living in $
    \mathcal{H}^1 := \{\phi: ||\phi||_{L^2(\Omega_k)} + ||\nabla \phi||_{L^2(\Omega_k)} < \infty \}$.

\subsection{Numerical stabilization of ice-flow models}

At each time step in a simulation, it is standard practice to solve the SSA momentum equations \eqref{weak_no_boundaries} first, followed by the implicit Euler thickness evolution equation, (\ref{eq:Hweak}). This means that, by necessity, the $H ^2$ on the right-hand side of the SSA momentum equations \eqref{weak_no_boundaries} is evaluated at the time step $k$, not $k+1$. Due to this explicit handling of the driving force of the SSA momentum equations, numerical instabilities appear when too large a time-step size is used, which is the core issue addressed in this paper. A similar problem appears for Stokes problems in mantle convection simulations, for which Kaus et al., 2010 \cite{Kaus2010} introduced the free-surface stabilization algorithm (FSSA). The FSSA modifies the weak form of the Stokes momentum equations, $ \int_{\Lambda} (\nabla \cdot \boldsymbol{\sigma}) \cdot \mathbf{w} ~d\Lambda= \int_{\Lambda}\rho_i \mathbf{g}\cdot \mathbf{w} ~d\Lambda,$ so that the driving stress (right hand side) is approximately evaluated at time-step $k+1$ instead of $k$ to account for the expected ice surface evolution.

The driving stress depends on time due to the time-dependence of the thickness $H$. In the p-Stokes case, the $H$-dependence is rather implicit, as it is due to the integration domain, $\Lambda$(x,y,z), being dependent on $H$. An approximation of the integration domain at the next time-step, $\Lambda_{k+1}$, can be found by an explicit Euler discretization of the Reynold's transport theorem,
\begin{align} \label{eq:stokes_FSSA}
   \int_{\Lambda_{k+1}}\rho_i\mathbf{g} \cdot \mathbf{w} ~d\Omega&\approx \int_{\Lambda_{k}}\rho_i\mathbf{g} \cdot \mathbf{w} ~d\Omega+ 
    \Delta t\int_{\partial \Lambda_k} (\mathbf{u \cdot \hat{\mathbf{n}}}) (\rho_i\mathbf{g} \cdot \mathbf{w})~d\Gamma.
\end{align}
where the last term represents the adjustment in the ice load given the predicted displacement of the surface elevation between the current and the next time step (see e.g. \citep{Loefgren2022,Loefgren2024}). In this formulation, the right-hand side is predicted at the next time step. When coupled to a free-surface evolution equation, this modification treats the driving stress implicitly. FSSA was later adapted to a grounded ice-sheet model by \cite{Loefgren2022}, also incorporating glaciological processes such as accumulation and ablation. The approach increases numerical stability of p-Stokes ice sheet models significantly \citep{Loefgren2022,Loefgren2024}. For marine-ice sheet simulations, Durand et al., 2009 \cite{Durand2009} introduced the sea spring stabilization scheme in the p-Stokes equations to predict the ocean pressure at the next time step. 
 
 \subsubsection{The thickness stabilization scheme (TSS)}\label{TSS}

The FSSA cannot directly be applied to stabilize SSA models. Firstly, this is because for SSA models, the integration domain $\Omega(x,y,t)$ is $H$-independent, but the right hand side driving stress is instead $H$-dependent due to the $H^2$-term on the integrand. Secondly, the FSSA is tailored to Stokes flow with a free surface, where all momentum equation variables, $(u_x,u_y,u_z,p)$, are solved for. In contrast, the SSA assumes plug flow and depth-independent horizontal velocities, rendering the FSSA framework unsuitable due to its reliance on a vertical velocity component. In this study, we develop a numerical stabilization scheme specifically for the SSA called the Thickness Stabilization Scheme (TSS), which does not rely on a full 3D velocity vector. Like FSSA, TSS modifies the right-hand side of the momentum equations by incorporating a prediction of the driving stress at the next time step, thereby also mimicking an implicit scheme.

This is done by replacing the squared ice thickness at the current time step, $H_k^2$, with the predicted thickness at the next time step, which we denote as $\tilde{H}^2_{k+1}$. This leads to the modified variational problem,
\begin{equation}
    \int_{\Omega}\mathbf{T} : \nabla \mathbf{v} ~d\Omega = \rho' \int_{\Omega} \tilde{H}_{k+1}^2 \nabla\cdot \mathbf{v} ~d\Omega.
\end{equation}
In order to find an expression for $\tilde{H}_{k+1}^2$, the thickness evolution equation,
\begin{align}
 \frac{\partial H_k}{\partial t} = -\nabla \cdot (H \mathbf{u}) + a_s - a_b ,
\end{align}
is adapted by multiplying by $2 H$ so that
\begin{align}
 \frac{\partial H^2}{\partial t} = 2H \frac{\partial H}{\partial t} =
    &= 2H (-\nabla \cdot (H \mathbf{u}) + a_s - a_b ).
\end{align}
Discretizing in time using a forward Euler scheme \eqref{eq:forwardH} results in
\begin{align} \label{eq:predH}
    \tilde{H}^2_{k+1} = H^2_k - 2 \Delta t H_k \nabla \cdot (H_k \mathbf{u}) + 2 \Delta t H (a_s - a_b).
\end{align}
Inserting the above expression into \eqref{weak_no_boundaries} yields the TSS-stabilized SSA equations,
\begin{equation}
\begin{aligned}
    \int_{\Omega} \mathbf{T} : \nabla \mathbf{v} ~d\Omega &+ \int_{\Omega} 2 \theta \Delta t H_k \nabla \cdot (H_k \mathbf{u})\nabla\cdot \mathbf{v} ~d\Omega \\
    &= \rho' \int_{\Omega} \big[H^2_k  - 2 \theta \Delta t H_k (a_b - a_s )\big] \nabla\cdot \mathbf{v} ~d\Omega.
\end{aligned}
\end{equation}
Here, a stabilization parameter, $\theta \in [0,1]$, has been introduced, allowing the user to choose between using no stabilization $(\theta = 0)$ and using the numerical scheme $(\theta = 1)$. Note that the TSS is constructed with an explicit Euler time discretization and is inserted into the SSA formulation, whereas the geometry evolution is solved using an implicit Euler time discretization, as in (\ref{eq:Hweak}). 

\subsubsection{Artificial viscosity}\label{arti}

To handle spurious oscillations, a simple artificial viscosity term is added to the thickness evolution equation \cite{VonNeumann1950},
\begin{align} \label{eq:Hweak_art}
    \int_{\Omega} [ H_{k+1} &+ \Delta t \nabla \cdot (H_{k+1}\mathbf{u}) ] \phi \, d\Omega - \int_{\Omega}\Delta t \mu_\mathrm{art} \nabla H_{k+1} \cdot \nabla \phi \, d\Omega \\
    &= \int_{\Omega} [ H_k + \Delta t (a_s - a_b )] \phi \, d\Omega,
\end{align}
where the artificial viscosity is defined as
\begin{equation}\label{muart}
    \mu_{\text{art}} = \alpha h |\mathbf{u}|.
\end{equation}
Here, $h$ is the local mesh size, $\alpha$ is a user-defined parameter, and $|\mathbf{u}|$ is the horizontal velocity magnitude. In this study, we use a default value of $\alpha = 0.1$. The use of an artificial viscosity term has a history in ice-sheet modeling (e.g. \citep{MacAyeal1997}), though a variety of more accurate approaches are also available. In this study, it serves as a practical choice, although care should be taken to ensure accuracy in simulations \citep{Wirbel2020, dosSantos2021}.

\section{Numerical experiments}
\subsection{Model implementation} \label{sq:setup}

In order to investigate the applicability of the TSS, the SSA equations were solved using \texttt{FEniCS}, an open-source software for solving PDEs using the finite element method \cite{FenicsSite, FenicsBook}. The aim of these simulations is to test whether the TSS increases numerical stability, i.e. whether the largest stable time step increases in TSS-stabilized simulations compared to simulations without TSS, and also how the TSS impacts accuracy and efficiency.

\begin{figure}
    \centering
    \includegraphics[width=0.5\linewidth]{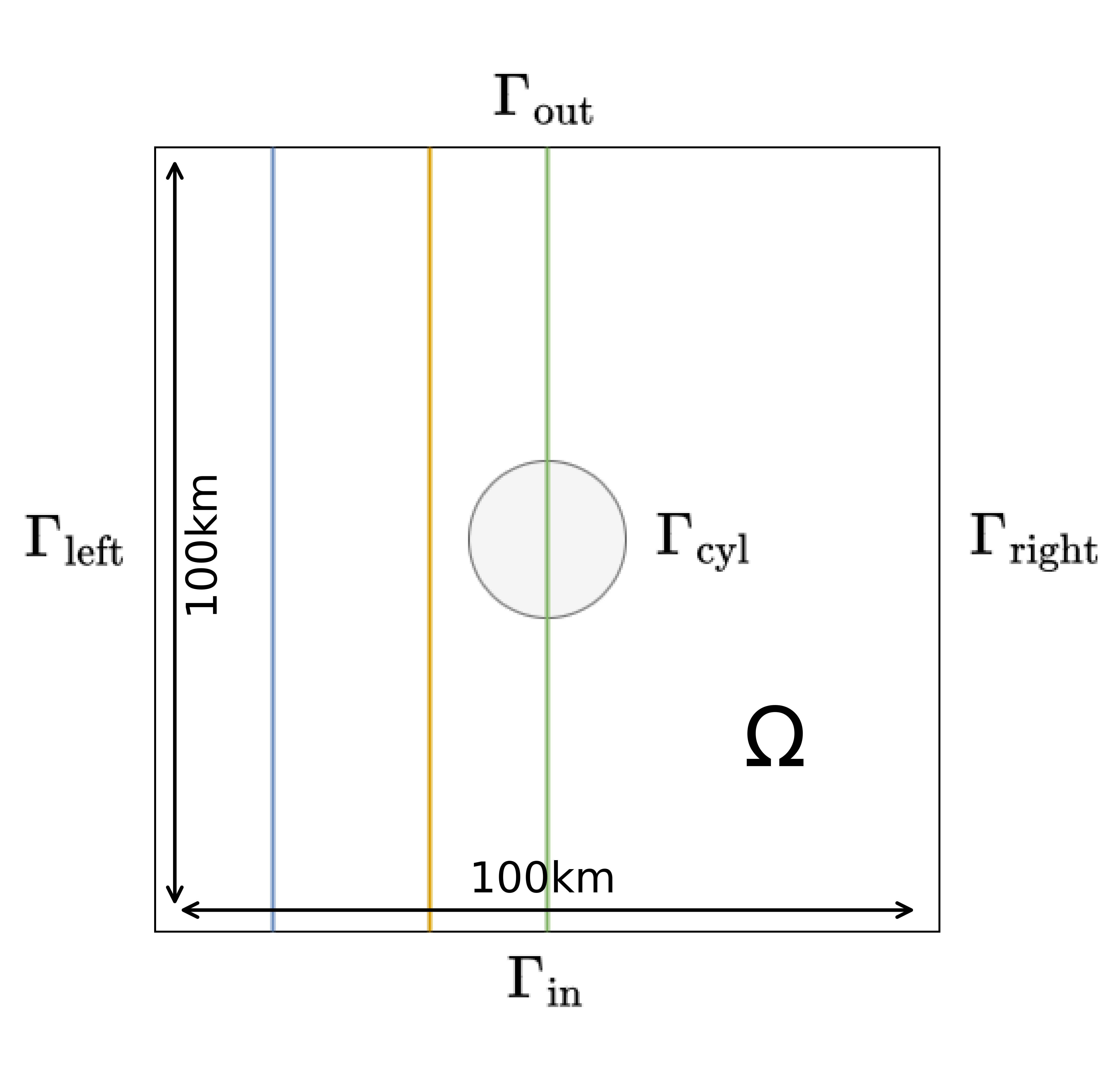}
    \caption{A schematic of the 2D model domain, $\Omega$, used in the simulations. The square domain represents a floating ice shelf, with boundaries indicated by $\Gamma_\mathrm{in}$ (inflow), $\Gamma_\mathrm{out}$ (calving front), $\Gamma_\mathrm{left}$ and $\Gamma_\mathrm{right}$ (lateral sides), and $\Gamma_\mathrm{cyl}$ (cylindrical obstacle). Cross sections for simulation analysis at $x=15$ km (blue), $x=35$ km (yellow), and $x=50$ km (green) are indicated.}
    \label{fig:domain}
\end{figure}

\subsubsection{Initial conditions and boundary conditions}

The SSA equations are solved on a square $100 \times 100$ km domain, $\Omega$, with a boundary,$\partial \Omega = \Gamma_\mathrm{in} \cup \Gamma_\mathrm{out} \cup \Gamma_\mathrm{left} \cup \Gamma_\mathrm{right} \cup \Gamma_\mathrm{cyl}$ (Figure \ref{fig:domain}). The inflow and outflow boundaries are denoted by $\Gamma_\mathrm{in}$ and $\Gamma_\mathrm{out}$, respectively. In this study, a constant normal velocity is set at the inflow boundary, $\Gamma_\mathrm{in}$. In order to represent the calving front or outflow boundary, $\Gamma_\mathrm{out}$, where the ice breaks off into the open ocean, a balance of forces is applied so that the ocean pressure balances with the driving stress. Making the assumption that, in reality, the ice would flow predominantly from the inflow boundary to the calving front, we apply an impenetrability condition on the lateral boundaries, $\Gamma_\mathrm{left}$ and $\Gamma_\mathrm{right}$. At the lateral boundaries of the domain, impenetrability boundary conditions are applied, so that $u_x = 0$ on the left and right-hand side boundaries. The ice initially has a uniform thickness, $H_0$, and is kept fixed at that value through time only at the inflow boundary.
We model the flow around an idealized, circular obstacle, which represents a pinning point in an ice shelf \cite{Favier2015, Henry2022}. Pinning points occur where otherwise floating ice is locally grounded on elevated bedrock. The circular obstacle with a radius of 10~km is placed in the center of the domain, located at $(x, y) = (50, 50)$~km (Figure~\ref{fig:domain}). An impenetrability condition is applied at the cylinder boundary, so that $\mathbf{u} \cdot \hat{\mathbf{n}} = 0$.

\subsubsection{Numerical experiments}

The problem is discretized with unstructured finite elements of degree one for both the velocity and the thickness. To investigate the accuracy and efficiency of the TSS, a number of simulations are performed with fixed parameters listed in Table~\ref{tab:parameters} in the Appendix, with and without TSS for $2\,000$ years. Simulations are performed with combinations of (1) an initial ice thickness of $H = 300$\,m and inflow velocity of $u_{y,0} = 300$\,m/yr, which we name the \textit{low-shear} experiment, and (2) an initial ice thickness of $H_0 = 1\,000$\,m and inflow velocity of $u_{y,0} = 1\,000$\,m/yr, which we name the \textit{high-shear} experiment (Tables~\ref{tab:simulations} in the Appendix). The low-shear simulations are performed with time-step sizes of $\Delta t = 0.5, 10, 40$, $50$ and $100$, and the high-shear simulations are performed with time-step sizes of $\Delta t = 0.5, 10$ and $100$. Each of these simulations was performed with ($\theta = 1$) and without ($\theta = 0$) TSS. The results were compared to reference solutions ($\Delta t = 0.5, \theta = 0$).

Further numerical experiments were performed for simulations with TSS ($\theta = 1$) for time-step sizes of $\Delta t = 1\,000$, and $10\,000$ years with an initial ice thickness of $H_0 = 300$\,m and a normal inflow velocity of $300$\,m/yr (Table~\ref{tab:parameters} in the Appendix). Due to the computational expense of these longer simulations, results are compared to a reference simulation of $\Delta t = 40$ without TSS ($\theta = 0$).

\begin{table}[h!]
\centering
\caption{Model parameters used in the simulations.}
\begin{tabular}{llll}
\hline
Parameter & Value & Units & Description \\
\hline
$A$ & $4.6 \times 10^{-25}$ & Pa\,m$^{-1/3}$\,s & Ice fluidity \\
$\rho_i$ & $900$ & kg\,m$^{-3}$ & Ice density \\
$\rho_w$ & $1\,000$ & kg\,m$^{-3}$ & Ocean density \\
$g$ & $9.8$ & m\,s$^{-1}$ & Gravity \\
$n$ & $3$ & - & Glen’s flow law exponent \\
\hline
\end{tabular}
\label{tab:parameters}
\end{table}

\subsubsection{Error and convergence estimation}

To calculate the deviation between a computed field and a reference solution, an error based on the average absolute difference over the domain is used. For a quantity $h$ (such as the ice thickness or the absolute velocity), the error is defined as
\begin{align}
    \epsilon_h = \frac{1}{|\Omega|} \int_{\Omega} \left| h - h^* \right| \, d\Omega, \label{eq:mae}
\end{align}
where $h^*$ denotes the reference solution and $|\Omega|$ is the area of the domain.

The convergence behavior of the solution over time is calculated via an update norm, which measures how much a field changes between time steps. For a field $h$, the update norm from time step $k$ to $k+1$ is defined as  
\begin{align}
    \delta_k =  \left\| h_{k+1} - h_k \right\|_{L^2(\Omega)}, \label{eq:update_norm}
\end{align}
where $\| \cdot \|_{L^2(\Omega)}$ denotes the $L^2$-norm. As the solution approaches a steady state, the update norm indicates stability if it tends to zero.

\section{Results}

\begin{figure}
    \centering
    \includegraphics[width=0.9\textwidth]{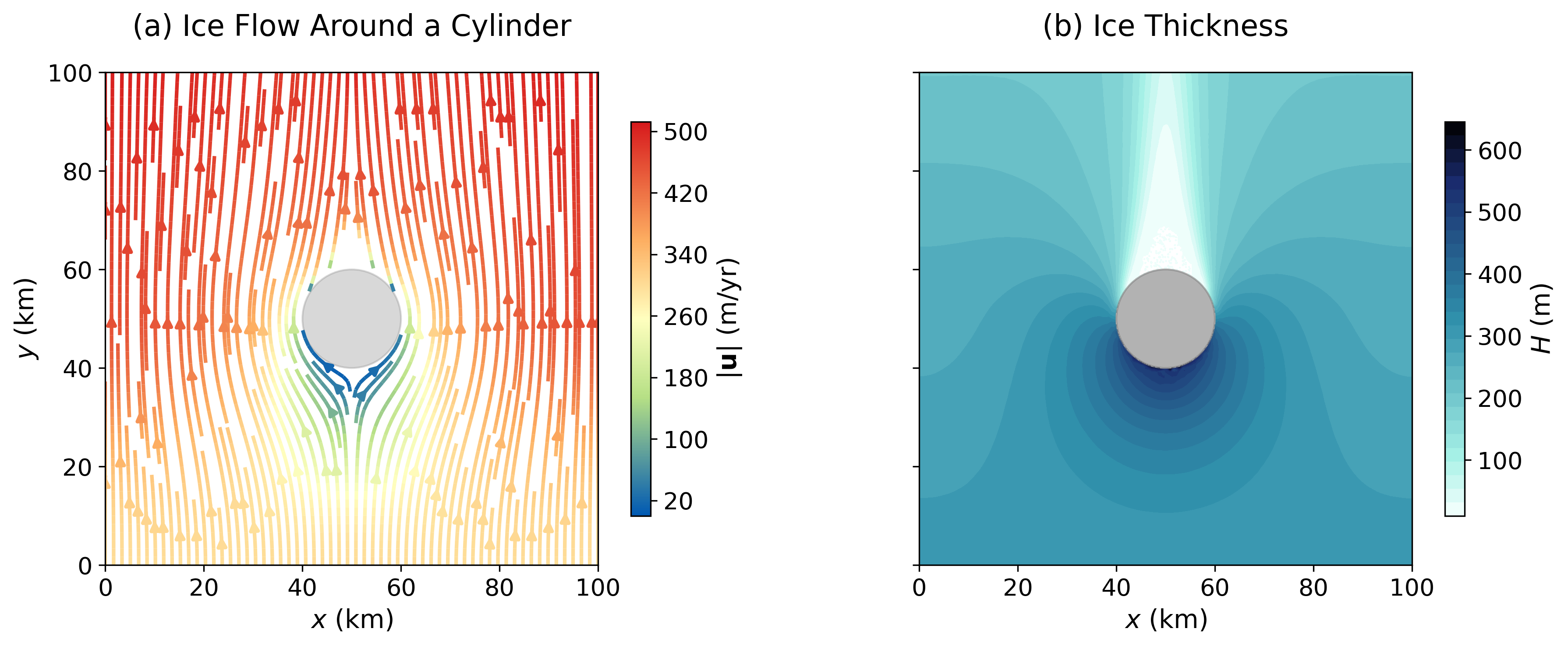}
    \caption{The $H_0 = 300$\,m, $u_{y,0} = 300$\,m/yr reference simulation at $2\,000$ years with a time-step size of $\Delta t = 0.5$~yrs and no TSS ($\theta = 0$). The panels show (a) the velocity streamlines around the obstacle with the color showing the velocity magnitude, and (b) the ice thickness field, $H$.}
    \label{fig:reference}
\end{figure}

During the simulation, both the ice thickness and velocity adjust due to the presence of the obstacle at the center of the domain as well as the boundary conditions imposed. On the upstream side of the obstacle, the ice slows down and thickens due to horizontally compressive stresses (Figs.~\ref{fig:reference}a). The presence of the obstacle causes the flow to be directed laterally around its sides, generating shear zones with high velocity gradients. On the lee side, downstream of the obstacle, the ice flow accelerates and thins as it flows towards the open ocean.

\subsection{Application of the Thickness Stabilization Scheme}

\subsubsection{Low-shear simulations}

\begin{figure}
    \centering
    \includegraphics[width=0.9\linewidth]{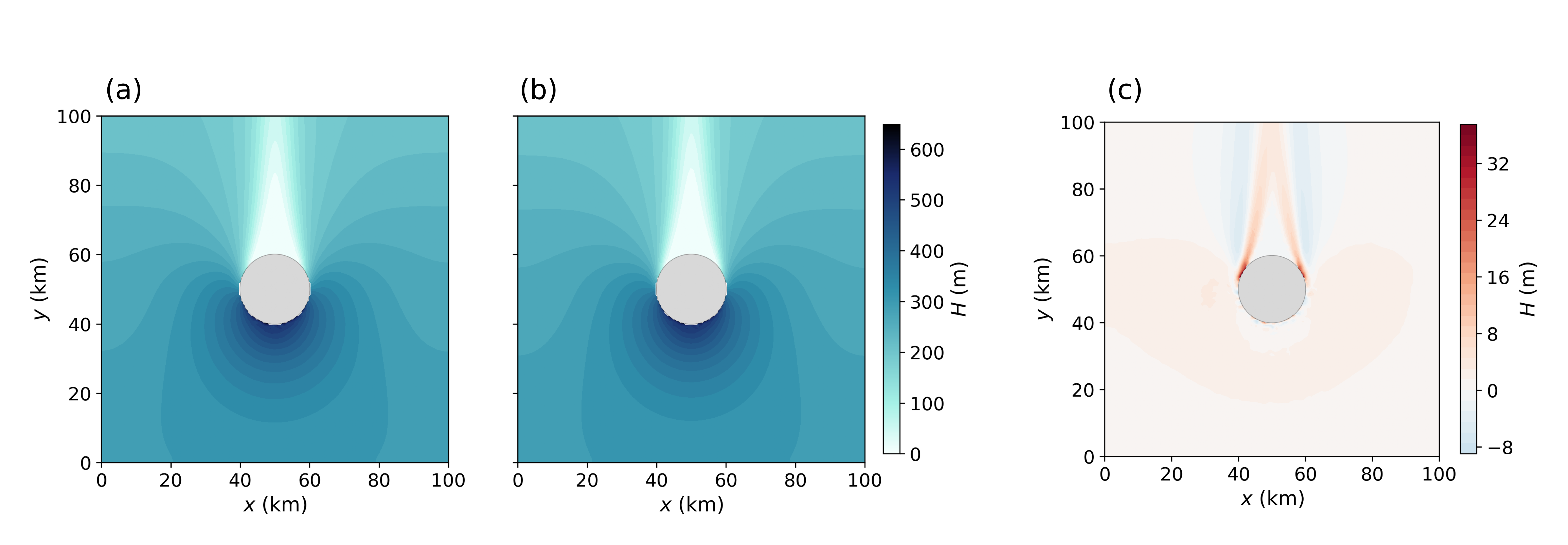}
    \caption{The final ice thickness, $H$, after $2\,000$ years of simulation with a time-step size $\Delta t = 40$ years for the $H_0=300$\,m, $u_{y,0}=300$\,m/yr initial conditions. The panels show (a) the ice thickness with TSS ($\theta = 1$), (b) the ice thickness without TSS ($\theta = 0$), and (c) the difference between the simulation with TSS and the simulation without TSS. The gray circle centered at $(x, y) = (50, 50)$ km indicates the location of the obstacle.}
    \label{fig:heatmapH_comparison_dt40}
\end{figure}

\begin{figure}
    \centering
    \includegraphics[width=0.9\linewidth]{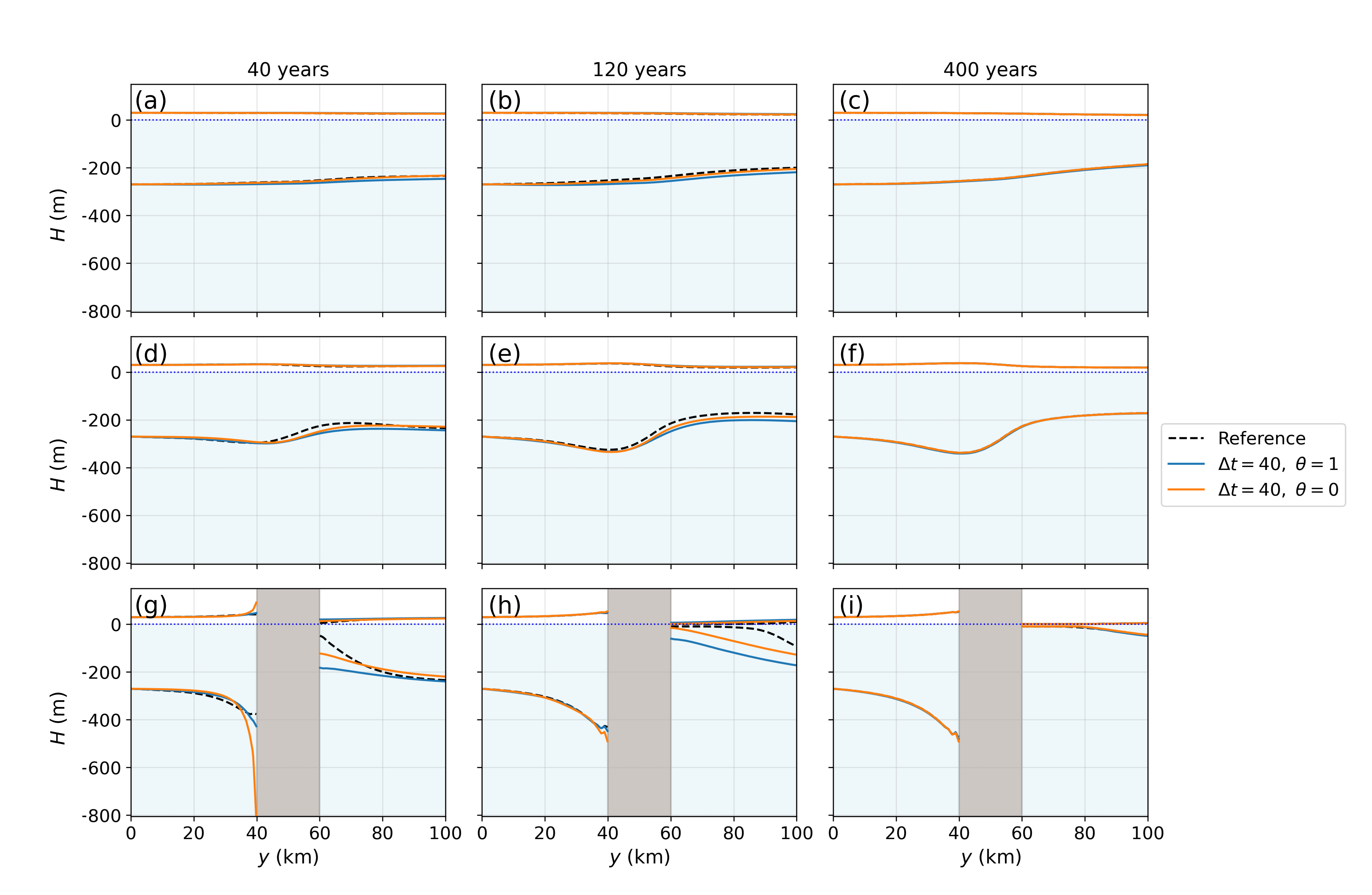}
    \caption{The upper ice surface, $z_s$, and the lower ice surface, $z_b$, at cross sections located at $x = 15~\mathrm{m}$, $x = 35~\mathrm{m}$, and $x = 50~\mathrm{m}$, evaluated after 40, 120, and 400 years for a time step of $\Delta t = 40$ years. The simulations are initialized with an ice thickness of $H_0 = 300$\,m and have an normal inflow velocity of $300$\,m/yr throughout the simulation (the low-shear scenario). Each panel compares the solution with TSS (blue), without TSS (orange), and the reference solution (dashed black). Shaded regions between $y = 40$ km and $y = 60$ km indicate the location of the obstacle.}
    \label{fig:HcrossTSScom}
\end{figure}

After simulating for $2\,000$ years with a time step of $\Delta t = 40$, the final ice thickness fields are nearly identical with and without TSS, with a maximum difference of approximately $40$\,m (Figure \ref{fig:domain}). The largest deviations are observed in the shear margins and on the lee side of the obstacle (Figs.~\ref{fig:heatmapH_comparison_dt40} and \ref{fig:HcrossTSScom}). In the shear zones, and in particular on the lee side of the obstacle, both simulations deviate most from the reference. After $40$ years, the simulation without TSS ($\theta = 0$) is slightly closer to the reference than the simulation with TSS ($\theta = 1$) across all cross sections. As time progresses to $120$ years, the simulations converge towards the reference, with the solution without TSS remaining closer. By $400$ years, the difference between the two simulations has decreased significantly and both show good agreement with the reference.

\begin{figure}
    \centering
    \includegraphics[width=0.9\linewidth]{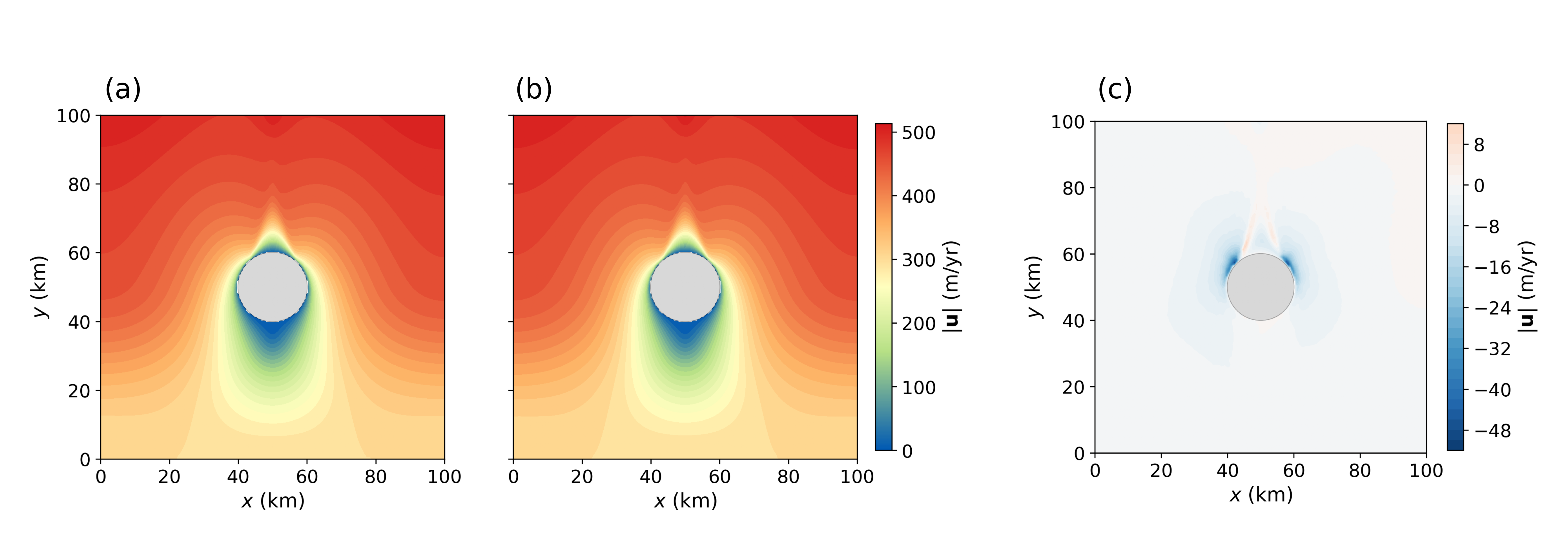}
    \caption{The absolute velocity, $|\mathbf{u}|$(m/yr), after $2\,000$ years of simulation with a time-step size of $\Delta t = 40$ years for the $H_0=300$\,m, $u_{y,0}=300$\,m/yr initial conditions. The panels show (a) the velocity field with TSS ($\theta = 1$), (b) the velocity field without TSS ($\theta = 0$), (c) the difference between the simulation with TSS and the simulation without TSS. The gray circle centered at $(x, y) = (50, 50)$ km shows the location of the obstacle.}
    \label{fig:heatmap_u_compare}
\end{figure}

\begin{figure}
    \centering
    \includegraphics[width=0.9\linewidth]{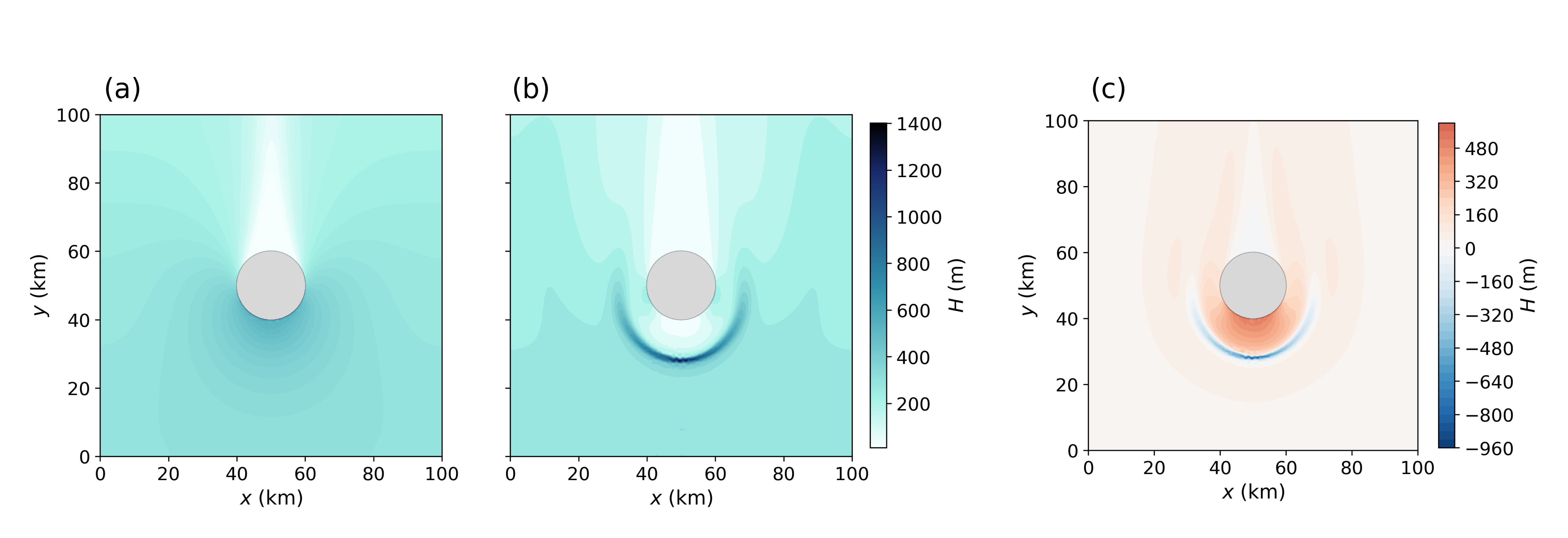}
    \caption{The ice thickness, $H$, after $2\,000$ years of simulation with a time-step size of $\Delta t = 50$ years for the low-shear scenario ($H_0 = 300, u_{y,0} = 300$). The panels show (a) the ice thickness with TSS ($\theta = 1$), (b) the ice thickness without TSS ($\theta = 0$), (c) the difference between simulation with TSS and the simulation without TSS. The obstacle is represented by the gray circle centered at $(50,50)$ km.}
    \label{fig:H50}
\end{figure}

When the time-step size is increased above $\Delta t = 40$, simulations without TSS ($\theta = 0$) begin to experience unphysical oscillations whereas simulations with TSS maintain a smooth solution. To showcase this, two simulations with a time-step size of $\Delta t=50$ are compared; one with TSS and the other without (Figure~\ref{fig:H50}). Notably, TSS allows time step sizes of $\Delta t = 100$ years to remain within 1\% accuracy, whereas time step sizes of $\Delta t = 45$ years produce significant error without TSS. This occurs on the stoss side of the obstacle, where a thickened band forms roughly $15$\,km upstream of the obstacle with very thin ice in between ($H\ll 100$\,m). Downstream of the obstacle, the ice is thin over a significantly larger portion of the domain.

This unphysical behavior is consistent with the results presented in Table~\ref{tab:errorsTSSonoff}, which show an increase in error as the time-step size, $\Delta t$, is increased for the ice thickness, $H$, and the absolute velocity, $|\mathbf{u}|$. For small time steps, both methods yield relatively low errors. For example, at $\Delta t = 10$, the error in $H$ is $0.1$ meters for the original SSA method $(\theta = 0)$ and $0.4$ meters for TSS $(\theta = 1)$. However, the error for the unstabilized method grows rapidly with an increasing time-step size. At $\Delta t = 45$, the error in $H$ increases to $57.4$ meters for $\theta = 0$, compared to just $1.5$ meters for $\theta = 1$. The velocity error follows a similar trend, rising from $0.1 ~\mathrm{m} / \mathrm{yr}$ to $63.1 ~\mathrm{m} / \mathrm{yr}$ for the unstabilized method, while the TSS solution remains accurate with an error of $1.5~\mathrm{m} / \mathrm{yr}$. For even larger time steps such as $\Delta t = 100$, the simulation without TSS shows significant deviation, with the velocity error reaching $2659.0~\mathrm{m} / \mathrm{yr}$, while TSS still maintains a relatively low error of $2.6 ~\mathrm{m} / \mathrm{yr}$. 

\begin{table}[ht]
\centering
\begin{tabular}{l|cc|cc}
\toprule
& \multicolumn{2}{c|}{$H$} 
& \multicolumn{2}{c}{$|\mathbf{u}|$} \\
{$\Delta t$} & $\theta = 0$ & $\theta = 1$ & $\theta = 0$ & $\theta = 1$ \\
\hline
10 & 0.103 & 0.441 & 0.131 & 0.476 \\
40 & 0.229 & 1.378 & 0.270 & 1.392 \\
45 & 57.352 & 1.508 & 63.124 & 1.512 \\
50 & 51.474 & 1.632 & 41.728 & 1.625 \\
100 & 224.675 & 2.726 & 2659.016 & 2.621 \\
\bottomrule
\end{tabular}
\caption{The error after $2\,000$ years for simulations with TSS ($\theta = 1$) and without TSS ($\theta = 0$) for ice thickness, $H$, and the absolute velocity, $|\mathbf{u}|$, compared to the reference solution ($\Delta t = 0.5$\,yr, $\theta = 0$).}
\label{tab:errorsTSSonoff}
\end{table}

\begin{figure}
    \centering
    \includegraphics[width=0.9\linewidth]{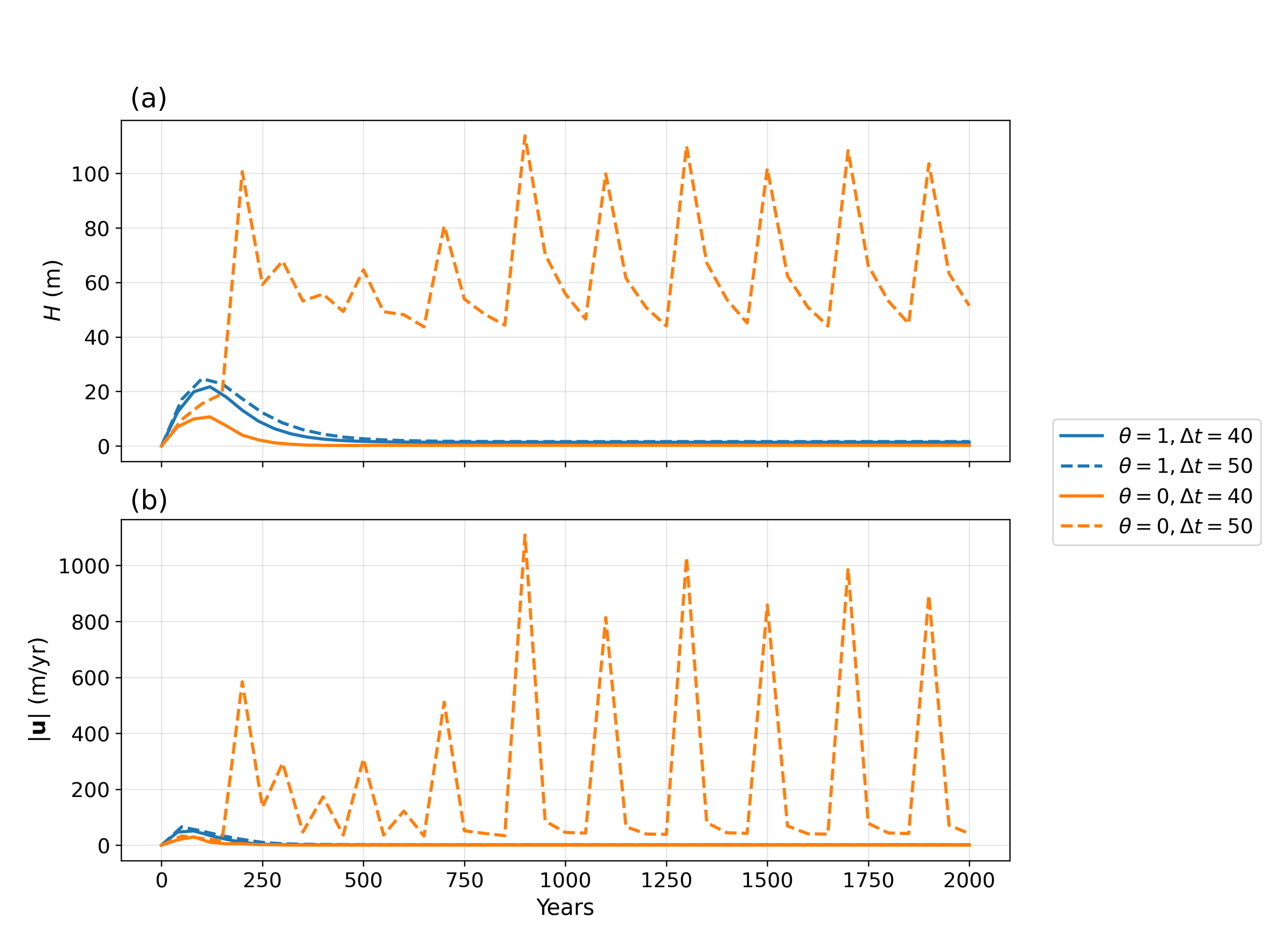}
    \caption{The domain-averaged absolute errors over time for simulations with an initial ice thickness of $H_0 = 300$\,m and a normal inflow velocity of $300$\,m/yr, measured relative to the reference simulation ($\Delta t = 0.5, \theta = 0$). The panels show (a) the mean  error in ice thickness, $H$(m), and (b) the mean error in the absolute velocity, $|\mathbf{u}|$(m/yr). The results are compared for simulations with TSS stabilization ($\theta = 1$) and without TSS ($\theta = 0$) at time-step sizes $\Delta t = 40$ and $50$ years. The reference solution uses $\Delta t = 0.5$ years and $\theta = 0$.}
    \label{fig:Average_H_u_error40}
\end{figure}

This trend in errors is also observed over time, as shown in Figure~\ref{fig:Average_H_u_error40}. The simulation without stabilization and with a time step of $\Delta t = 50$ shows oscillations, whereas all other configurations converge smoothly towards the reference solution.

\subsubsection{Runtime}
When performing simulations with and without TSS for $2\,000$ years, and different time-step sizes, the runtime is similar for both configurations. As seen in Figure~\ref{fig:runTSSonoff} in the Appendix, using a time-step size of $\Delta t = 40$ takes approximately $40$ seconds, which is $~2.5$ times faster than using a smaller time step of $\Delta t = 10$ for either simulation.

\subsubsection{TSS with larger time-step sizes}

Figure~\ref{fig:cross_sec_uy} in the Appendix shows the ice geometry at certain cross sections at $x = 35$~km and $x = 50$~km after $50~000$ years. The errors in ice thickness and velocity remain reasonable for time-step sizes of $\Delta t = 100$ and $\Delta t = 1\,000$ years, but errors are significantly larger for a time-step size of $\Delta t = 10\,000$ years, although these simulations remain numerically stable. Figure~\ref{fig:meanthick} in the Appendix shows the evolution of the ice thickness over time for time-step sizes of $\Delta t = 100$, $1\,000$ and $10\,000$ years, compared to the steady-state ice thickness of a simulation with a time-step size of $\Delta t = 40$ years without TSS. Within the first $2\,500$ years, the simulation with a time step of $\Delta t = 100$ years tends towards a steady-state mean ice thickness similar to the $\Delta t = 40, \theta = 0$ simulation. The simulation with a time-step size of $\Delta t = 1\,000$ years initially overshoots before also tending toward the $\Delta t = 40, \theta = 0$ solution. In contrast, the $\Delta t = 10~000$ year simulation does not tend towards the steady-state ice thickness of the $\Delta t = 40, \theta = 0$ simulation within $50\,000$ years.

\begin{table}
\centering
\vspace{0.2cm}
\begin{tabular}{c|cc|cc}
\toprule
$\Delta t$ & $\epsilon_H$ & $r_{H}$ & $\epsilon_{|\mathbf{u}|}$ & $r_{|\mathbf{u}|}$ \\
\hline
100 & 2.840 & 17.761 & 2.548 & -5.110 \\
1\,000 & 9.588 & 125.186 & 8.339 & 2.094 \\
10\,000 & 57.900 & 37.057 & 72.211 & -66.877 \\
\bottomrule
\end{tabular}
\vspace{0.3cm}
\caption{The mean error, $\epsilon$, and the range difference, $r = \mathrm{range}(\cdot) - \mathrm{range}_{\text{ref}}(\cdot)$, 
for the thickness, $H$, and the absolute velocity, $|\mathbf{u}|$, after a simulation time of 50~000 yrs, compared to a reference simulation 
($\Delta t = 40, \theta=0$). The error was computed using Equation~(\ref{eq:mae}).}
\label{tab:TSS_accuracy}
\end{table}

\subsubsection{High-shear simulations}

\begin{figure}
    \centering
    \includegraphics[width=0.9\linewidth]{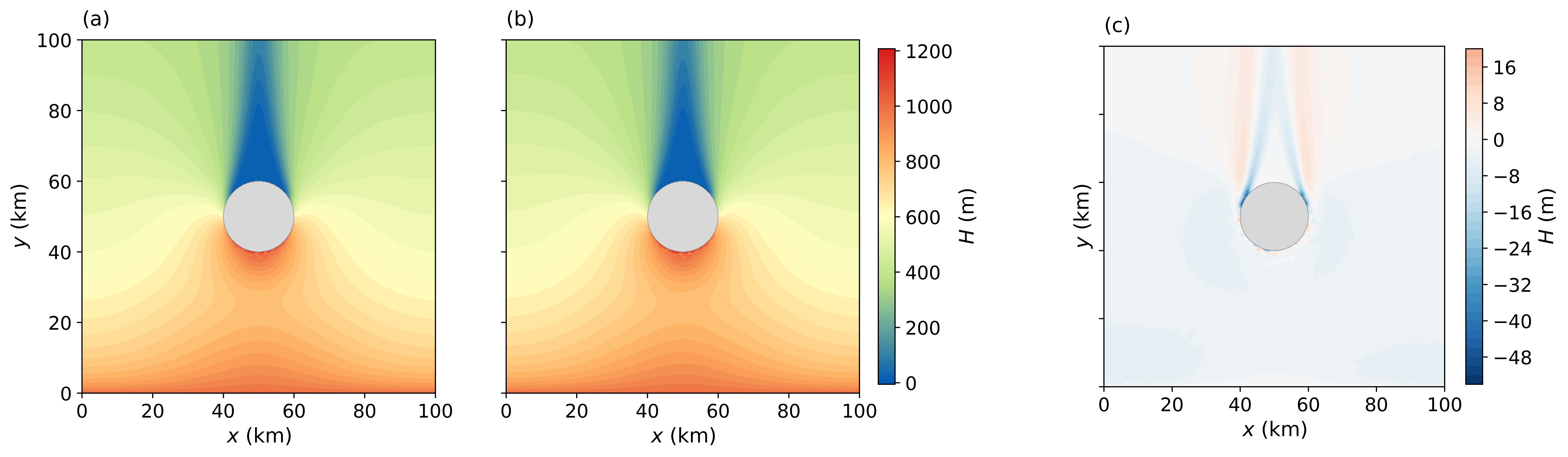}
    \caption{The ice thickness, $H$, after $2\,000$ years of simulation with a time-step size of $\Delta t = 5$ years for the high-shear scenario ($H_0 = 1\,000, u_{y,0} = 1\,000$). (a) The ice thickness with TSS ($\theta = 1$), (b) The ice thickness without TSS ($\theta = 0$), (c) The difference between (a) and (b). The obstacle is represented by the gray circle centered at $(50,50)$ km.}
    \label{fig:H5}
\end{figure}

\begin{figure}
    \centering
    \includegraphics[width=0.9\linewidth]{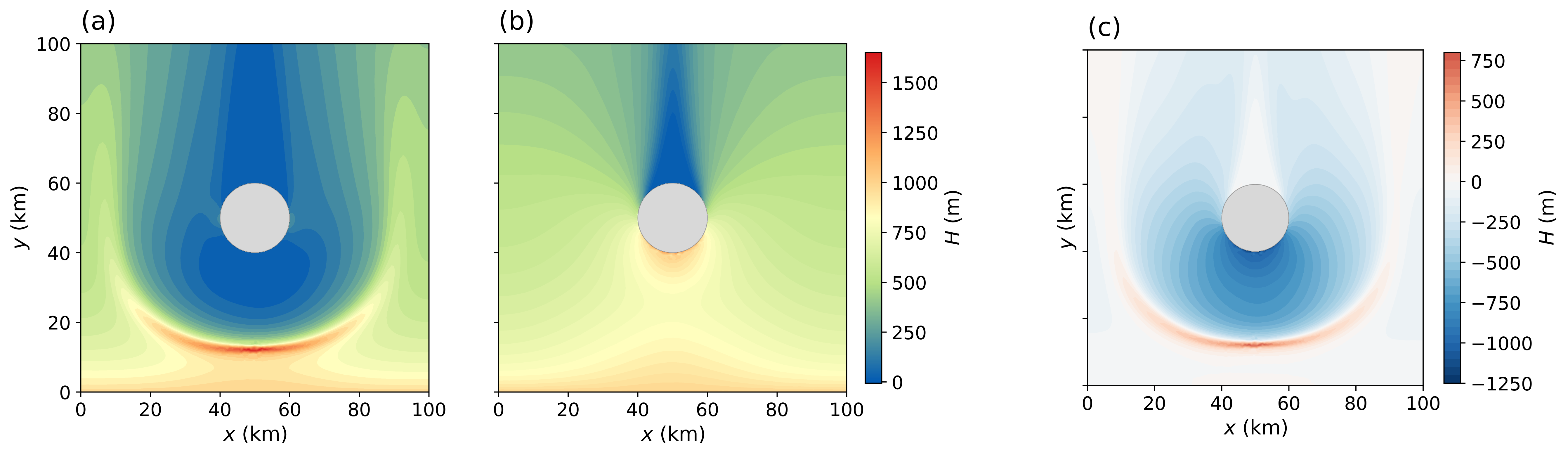}
    \caption{The ice thickness, $H$, after $2\,000$ years of simulation with a time-step size of $\Delta t = 10$ years for the high-shear scenario ($H_0 = 1\,000, u_{y,0} = 1\,000$). (a) The ice thickness with TSS ($\theta = 1$), (b) The ice thickness without TSS ($\theta = 0$), (c) The difference between (a) and (b). The obstacle is represented by the gray circle centered at $(50,50)$ km.}
    \label{fig:H10}
\end{figure}

Table~\ref{tab:HighShear} shows the mean error in ice thickness and velocity for the high-shear simulations, with an initial ice thickness of $H_0=1\,000$\,m and an inflow velocity of $u_{y,0} = 1\,000$\,m/yr, for time-step sizes of $\Delta t = 5$, $10$ and $100$ years. For time-steps sizes of $\Delta t = 5$, the error in ice thickness is small and slightly larger when including TSS (Figures~\ref{fig:H5} and \ref{fig:H10}). With time-step sizes of $\Delta t = 10$ and $\Delta t = 100$ years, the error in ice thickness is $42.5$ and $18.3$ times larger without TSS, respectively. The velocity error is $423.4$ and $12\,587.8$ times larger without TSS for time-step sizes of $\Delta t = 10$ and $\Delta t = 100$ years, respectively.

\begin{table}[ht] 
\centering
\begin{tabular}{l|cc|cc}
\toprule
& \multicolumn{2}{c|}{$H$} 
& \multicolumn{2}{c}{$|\mathbf{u}|$} \\
{$\Delta t$} & $\theta = 0$ & $\theta = 1$ & $\theta = 0$ & $\theta = 1$ \\
\hline
5 & 0.298 & 2.831 & 0.994 & 0.802 \\
10 & 208.181 & 4.895 & 5627.960 & 13.292 \\
100 & 554.254 & 30.370 & 91513.151 & 7.270 \\
\bottomrule
\end{tabular}
\caption{The error compared to the reference solution for the parameter choice of $H_0 = 1\,000$, $u_{y,0} = 1\,000$ (high-shear).}
\label{tab:HighShear}
\end{table}

\section{Discussion}

A similar stabilization scheme, called the free-surface stabilization algorithm (FSSA), was first implemented in Stokes—free-surface simulations in mantle convection \citep{Kaus2010}, and was later implemented in p-Stokes ice-sheet simulations at the ice-atmosphere interface \citep{Loefgren2022, Loefgren2024} and at the lower ice surface \citep{Henry2025FSSA}, enabling larger stable time steps to be taken. Unlike those studies, which rely on a full 3D velocity vector, the SSA equations do not solve for the vertical velocity component, making implementation of FSSA non-trivial. To address this, we developed the thickness stabilization scheme (TSS), which similarly treats certain terms implicitly and is suitable for the 2D, vertically-integrated SSA model. TSS has wider applicability in systems of coupled momentum and geometry evolution equations, and its efficiency could further be enhanced by adaptive time-stepping schemes, allowing small time steps during periods of rapid change and large time steps during near steady-state phases.

The TSS formulation presented here is limited to floating ice shelves and does not incorporate basal friction, which is critical in grounded ice dynamics. The SSA model is a depth-integrated approximation that neglects vertical shear, reducing accuracy when compared to higher-order models (e.g. p-Stokes models), in regions with pinning points \citep{Henry2022}. Furthermore, the simulations presented in this work were restricted to an idealized domain with a symmetric, circular obstacle that represents a pinning point \cite{Henry2022, Matsuoka2015, Reese2018}. Nevertheless, the approach presented here improves the efficiency of depth-integrated ice-flow models such as the SSA and is adaptable across model hierarchies for realistic, large-scale ice-sheet simulations.

\section{Conclusion}

In this study, a numerical stabilization method was developed for floating ice shelves using the Shallow Shelf Approximation (SSA). The goal was to enable larger stable time steps and and to improve numerical stability. The method, called the thickness stabilization scheme (TSS), works by mimicking an implicit treatment of the driving stress term, thereby dampening the rapid thickness changes that typically cause instability in explicit schemes. Without TSS, the solutions were stable and physically realistic only for time steps up to $\Delta t = 40$ years in simulations with low shear and $\Delta t = 5$ years with high shear. With TSS, simulations remained stable with time steps as large as $\Delta t = 10,000$ years, although such large steps led to unphysical behavior. Remarkably, using a time step of $\Delta t = 100$ year ensured high accuracy in low shear simulations, with errors remaining below 1\% for both the ice thickness and the velocity magnitude.

The ability to take larger time steps in ice-sheet simulations using SSA greatly reduces the computational cost, making TSS highly applicable for long-term ice-sheet modeling. The improved computational efficiency allows re-allocation of computational resources to increase, for example, the spatial resolution.. Furthermore, TSS has the potential to improve the computational efficiency of common large-scale coupled modeling frameworks that incorporate SSA. The implementation of the TSS in a depth-integrated model has the potential to be adapted to other reduced-order ice-sheet models such as Depth-Integrated Viscosity Approximation (DIVA, \cite{Goldberg2011}), the Blatter-Pattyn Approximation \cite{Blatter1995,Pattyn2003}, or as part of spatially-coupled frameworks such as the ISCAL method \cite{Ahlkrona2016}. Lastly, additional systems of equations that couple momentum equations to a mass conservation equation, such as the steady-state shallow-water equations, are ideal candidates for an analogous stabilization scheme.

\section{Acknowledgments}

ACJH is supported by the Wallenberg Foundation (KAW 2021.0275). JA was funded by the Swedish Research Council, grant number 2021-04001, as well as the Swedish e-Science Research Centre (SeRC). The authors acknowledge minimal assistance from ChatGPT (versions 4 \& 5, OpenAI) for code development. The authors thank André Löfgren for valuable discussions.

\section*{Author contribution}

ACJH conceptualized the Thickness Stabilization Scheme (TSS) in discussion with JA. ACJH derived the TSS for the Shallow Shelf Approximation and the model was implemented in FEniCS by ACJH and JA. ACJH designed the numerical experiments, which were performed and analysed by TWD, with input from ACJH. TWD and ACJH wrote the original draft, which was edited by all authors.

\newpage
\appendix
\section{Appendix}
\label{app1}

\begin{table}[h!]
\centering
\caption{List of simulations and their parameters.}
\begin{tabular}{lcccc}
\hline
$\theta$ & $\Delta t$ [a] & $H_0$ [m] & $u_{y,0}$ [m/yr] \\
\hline
0 & 0.5 & 300 & 300 & \\
0 & 10 & 300 & 300 & \\
0 & 40 & 300 & 300 & \\
0 & 45 & 300 & 300 & \\
0 & 50 & 300 & 300 & \\
1 & 0.5 & 300 & 300 & \\
1 & 10 & 300 & 300 & \\
1 & 40 & 300 & 300 & \\
1 & 45 & 300 & 300 & \\
1 & 50 & 300 & 300 & \\
\hline
0 & 0.5 & 1\,000 & 1\,000 & \\
0 & 5 & 1\,000 & 1\,000 & \\
0 & 10 & 1\,000 & 1\,000 & \\
0 & 100 & 1\,000 & 1\,000 & \\
1 & 0.5 & 1\,000 & 1\,000 & \\
1 & 5 & 1\,000 & 1\,000 & \\
1 & 10 & 1\,000 & 1\,000 & \\
1 & 100 & 1\,000 & 1\,000 & \\
\hline
1 & 100 & 300 & 300 & \\
1 & 1\,000 & 300 & 300 & \\
1 & 1\,0000 & 300 & 300 & \\
\hline
\end{tabular}
\label{tab:simulations}
\end{table}

\begin{figure}
    \centering
    \includegraphics[width=0.7\linewidth]{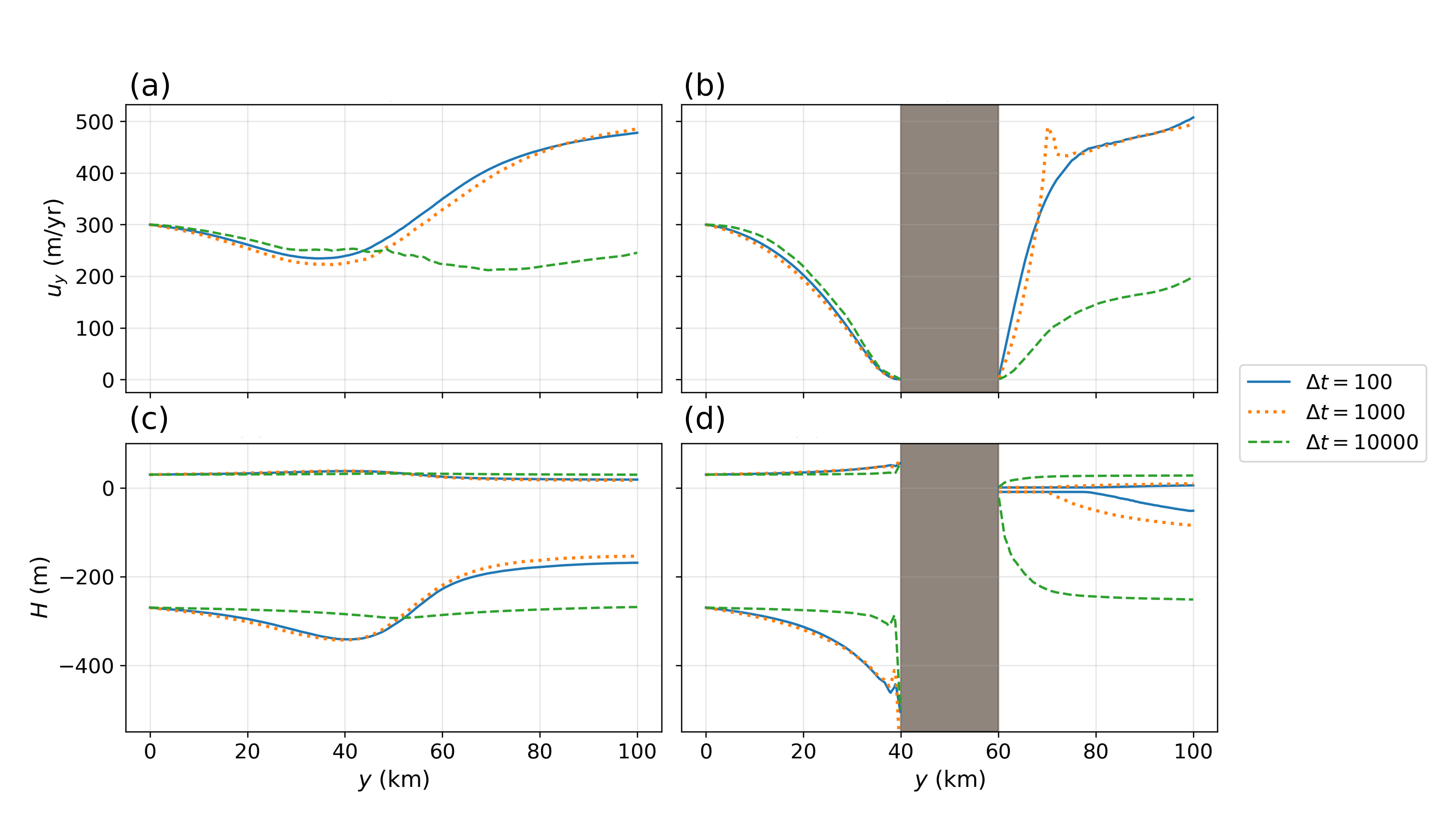}
    \caption{Cross sections at $x=35$ km and $x=50$ km of the thickness, $H$, and velocity, $u_y$, at $T = 50\,000$ years for simulations using TSS ($\theta = 1$) with different time-step sizes.}
    \label{fig:cross_sec_uy}
\end{figure}

\begin{figure}
    \centering
    \includegraphics[width=0.7\linewidth]{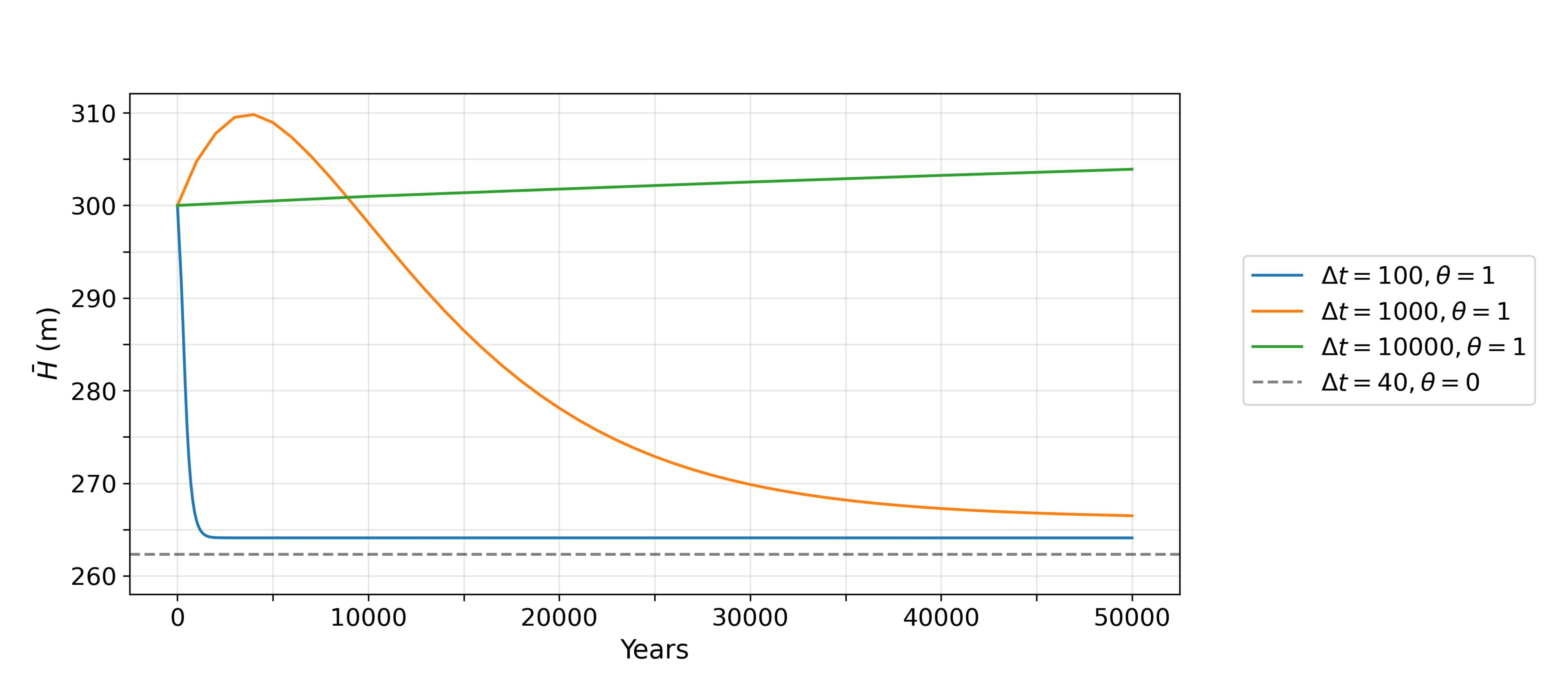}
    \caption{The mean thickness evolution over time for different time steps ($\Delta t$) using (TSS). The dashed line represents the reference mean asymptotic thickness value at $\Delta t = 40$, $\theta = 0$. The simulations were performed until $T=50\,000$ yrs.}
    \label{fig:meanthick}
\end{figure}

\begin{figure}
    \centering
    \includegraphics[width=0.5\linewidth]{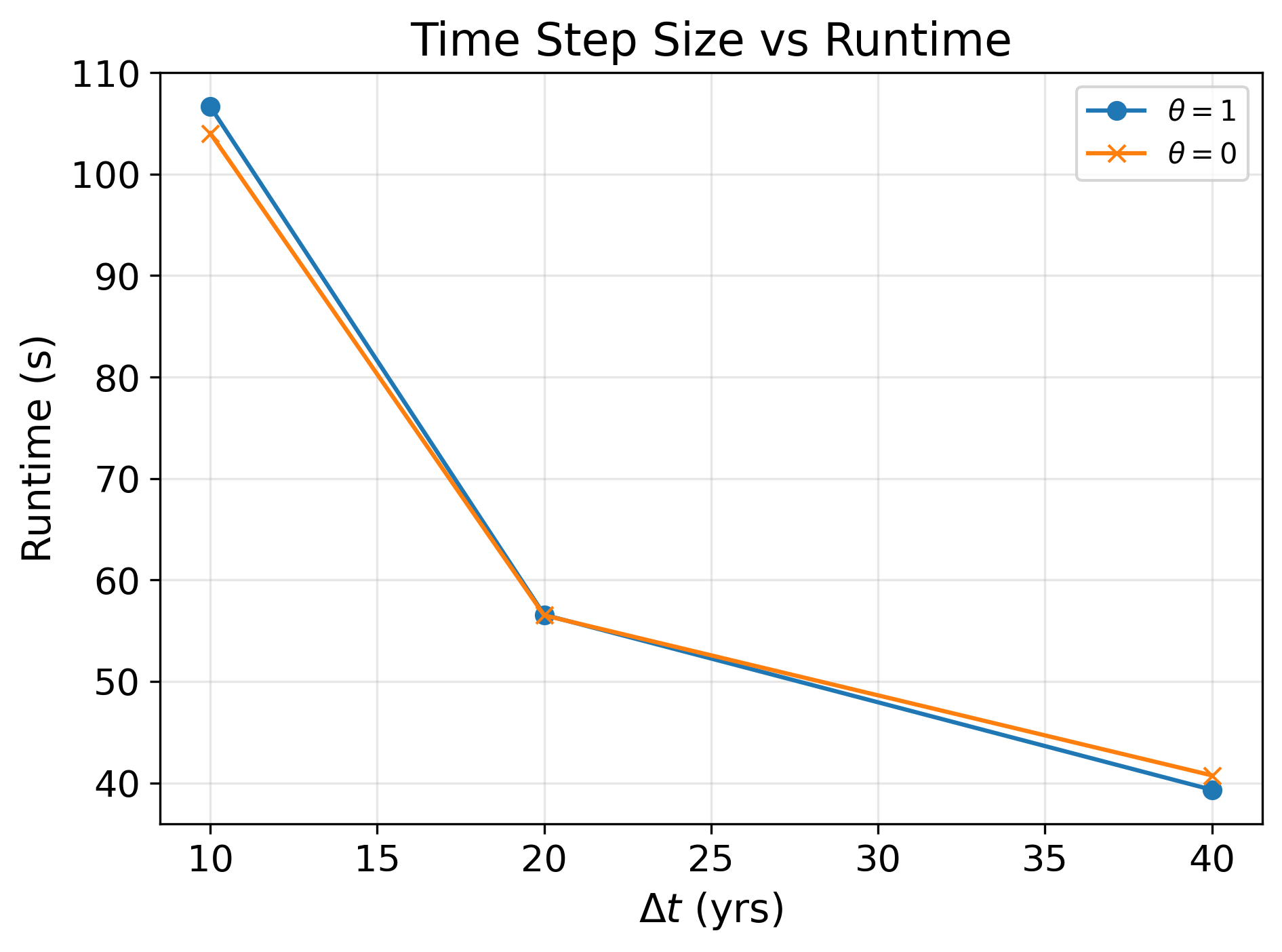}
    \caption{The CPU runtime for various time-step sizes, $\Delta t$, and both with TSS ($\theta = 1$) and without TSS ($\theta = 0$) for simulations performed over $2\,000$ years. These simulations are performed with an initial ice thickness of $300$\,m and a normal inflow velocity of $u_{y, 0} = 300$\,m/yr.}
    \label{fig:runTSSonoff}
\end{figure}

\begin{figure}
    \centering
    \includegraphics[width=0.5\linewidth]{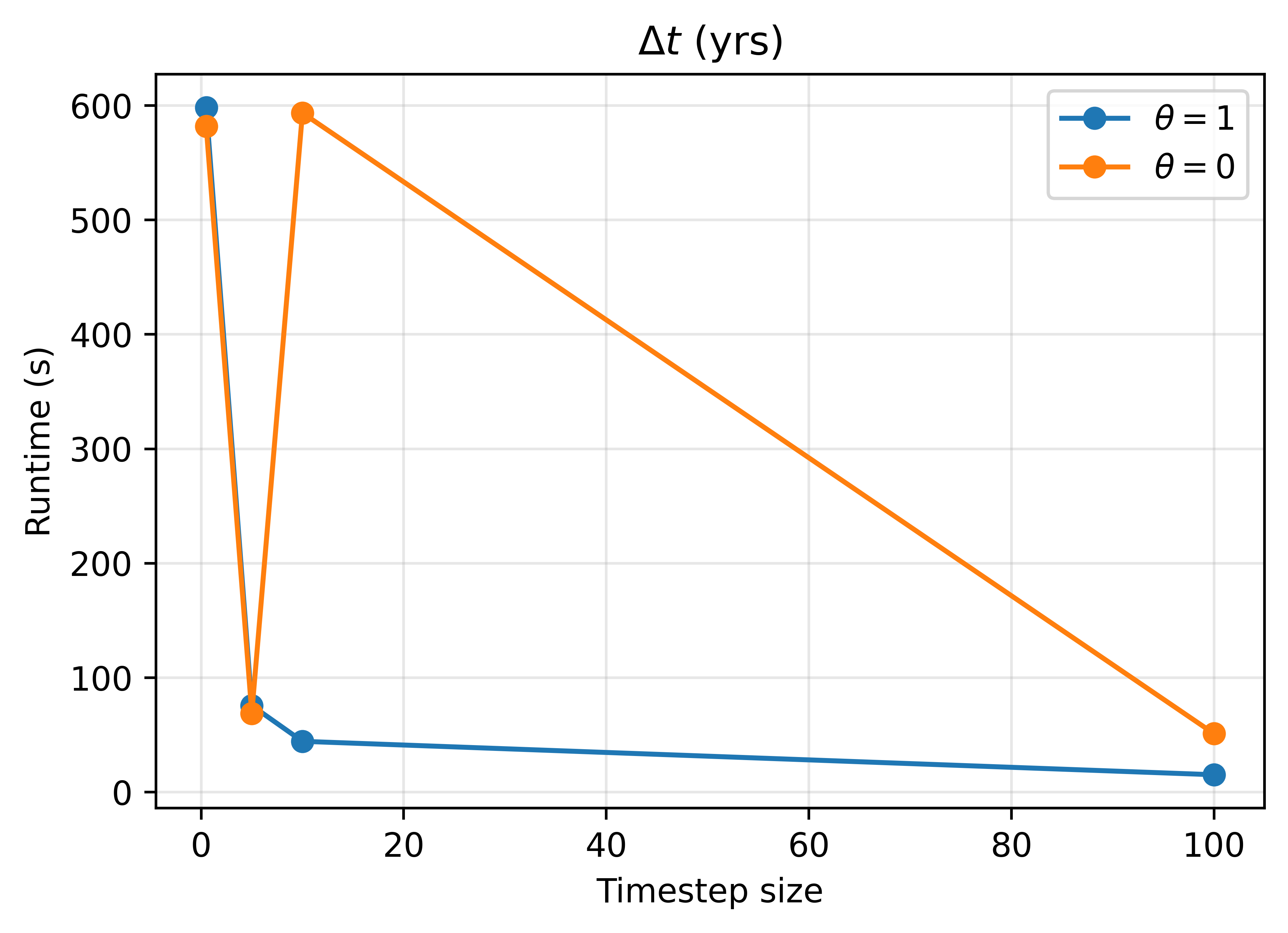}
    \caption{The CPU runtime for various time-step sizes, $\Delta t$, and both with TSS ($\theta = 1$) and without TSS ($\theta = 0$) for simulations performed over $2\,000$ years. These simulations are performed with an initial ice thickness of $1\,000$\,m and a normal inflow velocity of $u_{y, 0} = 1\,000$\,m/yr.}
    \label{Runtime1000}
\end{figure}

\newpage
\bibliography{main.bib}

\end{document}